# OBSERVATIONAL TEST OF STOCHASTIC HEATING IN LOW-β FAST SOLAR WIND STREAMS

Sofiane Bourouaine[1] & Benjamin D. G. Chandran
Space Science Center and Department of Physics, University of New Hampshire, Durham, NH 03824, USA
Draft version July 10, 2013

## ABSTRACT

Spacecraft measurements show that protons undergo substantial perpendicular heating during their transit from the Sun to the outer heliosphere. In this paper, we use *Helios 2* measurements to investigate whether stochastic heating by low-frequency turbulence is capable of explaining this perpendicular heating. We analyze *Helios 2* magnetic-field measurements in low-β fast-solar-wind streams between heliocentric distances $r = 0.29$ AU and $r = 0.64$ AU to determine the rms amplitude of the fluctuating magnetic field, $\delta B_p$, near the proton gyroradius scale $\rho_p$. We then evaluate the stochastic heating rate $Q_{\perp \rm stoch}$ using the measured value of $\delta B_p$ and a previously published analytical formula for $Q_{\perp \rm stoch}$. Using *Helios* measurements we estimate the 'empirical' perpendicular heating rate $Q_{\perp \rm emp} = (k_B/m_p) BV (d/dr)(T_{\perp p}/B)$ that is needed to explain the $T_{\perp p}$ profile. We find that $Q_{\perp \rm stoch} \sim Q_{\perp \rm emp}$, but only if a key dimensionless constant appearing in the formula for $Q_{\perp \rm stoch}$ lies within a certain range of values. This range is approximately the same throughout the radial interval that we analyze and is consistent with the results of numerical simulations of the stochastic heating of test particles in reduced magnetohydrodynamic turbulence. These results support the hypothesis that stochastic heating accounts for much of the perpendicular proton heating occurring in low-β fast-wind streams.

*Subject headings:* solar wind — turbulence — waves — MHD

## 1. INTRODUCTION

As solar-wind plasma flows away from the Sun, the temperatures of the different particle species decrease much more slowly than they would decrease if the solar wind were undergoing (double) adiabatic expansion, implying that the solar wind experiences some form of heating (Kohl et al. 1998; Marsch et al. 1982a,b). A number of observations and theoretical models suggest that this heating arises from the dissipation of Alfvén wave (AW) turbulence (e.g., Coleman 1968; Dmitruk et al. 2002; Cranmer et al. 2007; Verdini et al. 2010; Chandran et al. 2011). For example, AWs are observed in the corona (Tomczyk et al. 2007), and the energy flux carried by the observed AWs is sufficient to power the solar wind (De Pontieu et al. 2007). AW-like fluctuations are also seen in spacecraft measurements of the plasma velocity, electric field, and magnetic field in the interplanetary medium (Belcher & Davis 1971; Matthaeus & Goldstein 1982; Tu & Marsch 1995; Chen et al. 2011). The total turbulent heating rate inferred from these in situ measurements is sufficient to explain solar-wind temperature profiles (Smith et al. 2001; Vasquez et al. 2007; MacBride et al. 2008; Cranmer et al. 2009). In addition, solar wind models based on plasma heating by AW turbulence are consistent with Faraday rotation observations (Hollweg et al. 2010; see also Sakurai & Spangler 1994) and radio scintillation observations of the solar corona and solar wind (Harmon & Coles 2005; Chandran et al. 2009).

On the other hand, two observational results pose a difficult challenge for solar wind models based on heating by AW turbulence. First, UVCS observations show that O$^{+5}$ ions in coronal holes satisfy $T_\perp \gg T_\parallel$, where $T_\perp$ and $T_\parallel$ are perpendicular and parallel temperatures measuring thermal motions perpendicular and parallel to the magnetic field $B$ (Kohl et al. 1998; Li et al. 1998; Esser et al. 1999). Similarly, in situ measurements show that $T_\perp > T_\parallel$ within the core of the proton velocity distribution in low-β fast-solar-wind streams (Marsch et al. 1982b, 2004; Bourouaine et al. 2010), where $\beta = 8\pi p/B^2$, $p$ is the plasma pressure, and $B$ is the magnetic field strength. Second, the proton and electron temperature profiles measured by spacecraft at $r \geq 0.3$ AU imply that the proton heating rate exceeds the electron heating rate in the fast solar wind (Cranmer et al. 2009). These observations are difficult for AW turbulent heating to explain for the following reasons. The energy cascade in AW turbulence is anisotropic, transporting wave energy primarily from small $k_\perp$ to large $k_\perp$, and only weakly to larger $k_\parallel$, where $k_\perp$ and $k_\parallel$ are the components of the wave vector perpendicular and parallel to the background magnetic field (Shebalin et al. 1983; Goldreich & Sridhar 1995). Because the AW frequency is $\omega = k_\parallel v_A$, the absence of an efficient cascade to larger $k_\parallel$ implies that AW energy cascades only weakly to larger $\omega$. In particular, very little energy cascades to frequencies comparable to the ion cyclotron frequencies (Cranmer & van Ballegooijen 2003; Howes et al. 2008a). As a result, the AWs at small scales are unable to heat ions via a cyclotron resonance. On the other hand, when $k_\perp$ increases to values $\gtrsim \rho_p^{-1}$, where $\rho_p$ is the proton gyroradius, the AWs become kinetic Alfvén waves (KAWs). If the AWs/KAWs at large $k_\perp$ damp according to linear Vlasov/Maxwell theory, then the primary damping mechanism when $\beta \ll 1$ is electron Landau damping (Quataert 1998; Leamon et al. 1999; Gary & Nishimura 2004; Howes 2010), which causes parallel electron heating, rather than perpendicular ion heating.

These observations have motivated a number of studies of AW/KAW turbulent heating that go beyond the framework of linear Vlasov/Maxwell theory (e.g., Voitenko & Goossens 2004; Dmitruk et al. 2004; Markovskii et al. 2006; Bourouaine et al. 2008; Parashar et al. 2009; Lehe et al. 2009; Lynn et al. 2012; Wan et al. 2012; Wu et al. 2013; Karimabadi et al. 2013). In this paper, we focus on a nonlinear heating mechanism called "stochastic heating," which

[1] email: s.bourouaine@unh.edu

arises when fluctuating electric and/or magnetic fields at wavelengths comparable to a particle's gyroradius disrupt a particle's smooth gyromotion, leading to non-conservation of the particle's magnetic moment (McChesney et al. 1987; Karimabadi et al. 1994; Johnson & Cheng 2001; Chen et al. 2001; Chaston et al. 2004; Fiksel et al. 2009; Chandran 2010). Chandran et al. (2010) derived an analytic formula for the stochastic ion heating rate $Q_{\perp\text{stoch}}$ in low-$\beta$ plasmas as a function of the amplitude of AW/KAW fluctuations at the ion gyroradius scale. Our goal is to use this formula in conjunction with measurements from the *Helios* spacecraft to test the hypothesis that stochastic heating is responsible for the perpendicular ion heating that occurs in low-$\beta$ fast wind streams. In Section 2, we review previous work on the stochastic heating rate $Q_{\perp\text{stoch}}$. We also describe how the rate of perpendicular proton heating can be inferred from the radial profile of the proton temperature. We call this empirically determined heating rate $Q_{\perp\text{emp}}$. In Section 3.1, we present a detailed analysis using *Helios 2* measurements to evaluate $Q_{\perp\text{stoch}}$ and $Q_{\perp\text{emp}}$ at heliocentric distance $r = 0.29$ AU. By setting $Q_{\perp\text{stoch}} = Q_{\perp\text{emp}}$, we obtain a constraint on the two dimensionless constants appearing in the analytical formula for $Q_{\perp\text{stoch}}$ derived by Chandran et al. (2010). We then compare this constraint to the values of these constants that have been found in previous numerical studies of stochastic heating. In Section 3.2, we repeat this analysis at $r = 0.4$ AU and $r = 0.64$ AU. We discuss our results and conclusions in Section 4.

## 2. ANALYTICAL EXPRESSIONS FOR $Q_{\perp\text{STOCH}}$ AND $Q_{\perp\text{EMP}}$

If an ion moves in the presence of electric and magnetic fields that vary over a characteristic spatial scale $l$ and time scale $\tau$, and if the ion's gyroradius $\rho_i$ and cyclotron frequency $\Omega$ satisfy the inequalities $\rho_i \ll l$ and $\Omega\tau \gg 1$, then the ion's motion in the plane perpendicular to $B$ is nearly periodic. As a consequence, the ion's magnetic moment $\mu = mv_{\perp}^2/2B$, an adiabatic invariant, is almost exactly conserved (Kruskal 1962). Here, $m$ is the ion's mass, $v_{\perp}$ is the component of the ion's velocity perpendicular to $B$, $\Omega = qB/mc$, $q$ is the ion charge, and $\rho_i = v_{\perp}/\Omega$. On the other hand, if $l \sim \rho_i$, and if the amplitudes of the fluctuations in the electric and/or magnetic fields are sufficiently large, then the ion's motion in the plane perpendicular to $B$ ceases to be nearly periodic, even if $\Omega\tau \gg 1$, and magnetic moment conservation is violated (McChesney et al. 1987). In this case, the velocity-space average of the ion magnetic moment, $k_B T_{\perp}/B$, can increase in time. We refer to an increase in $k_B T_{\perp}/B$ as perpendicular heating. (An increase in $k_B T_{\perp}$ when magnetic moments are constant and $B$ increases is not really heating because it is reversible by subsequently reducing $B$.) Perpendicular heating resulting from the disruption of particle gyro-orbits by turbulent fluctuations with $\Omega\tau$ significantly greater than unity is called stochastic heating.

Most of the energy in solar wind turbulence is in the form of transverse, divergence-free fluctuations of the velocity and magnetic field (Goldstein et al. 1995; Tu & Marsch 1995). By "transverse," we mean that the fluctuating velocity and magnetic field vectors are preferentially aligned perpendicular to the background magnetic field (Goldstein et al. 1995). These properties are shared by one of the linear wave modes of the plasma at lengthscales greatly exceeding the proton gyroradius

$$\rho_p = v_{\perp p}/\Omega_p, \tag{1}$$

where

$$v_{\perp p} = \left(\frac{2k_B T_{\perp p}}{m_p}\right)^{1/2} \tag{2}$$

is the perpendicular thermal speed of protons, $T_{\perp p}$ ($T_{\parallel p}$) is the perpendicular (parallel) proton temperature, and $m_p$ is the proton mass. This wave mode is the Alfvén wave (AW). We thus refer to non-compressive, transverse fluctuations at scales $\gg \rho_p$ as AW turbulence. By this notation, we do not mean to infer that the fluctuations share all the properties of linear AWs. For example, we do not believe that the fluctuations in the solar wind vary monochromatically in time, since nonlinear interactions prevent this type of behavior. As mentioned in the introduction, the energy cascade in AW turbulence is anisotropic, transporting energy primarily to larger $k_{\perp}$, and only weakly to larger $k_{\parallel}$. As a result, at $k_{\perp}\rho_p \sim 1$ Alfvénic fluctuations are highly anisotropic, with $k_{\perp} \gg k_{\parallel}$, and have characteristic frequencies $\ll \Omega_p$, where $\Omega_p$ is the proton cyclotron frequency (Cranmer & van Ballegooijen 2003; Howes et al. 2008a). We refer to fluctuations with characteristic frequencies $\ll \Omega_p$ as "low-frequency" fluctuations. As $k_{\perp}\rho_p$ increases to values $\gtrsim 1$, the AW branch of the linear dispersion relation changes to the kinetic Alfvén wave (KAW) branch, and the AW cascade transitions to a KAW cascade (Schekochihin et al. 2009). We use the term AW/KAW turbulence to denote AW turbulence at scales $\gg \rho_p$ that transitions to KAW turbulence at scales $\lesssim \rho_p$.

Using phenomenological arguments, Chandran et al. (2010) derived an analytic formula for the stochastic heating rate (per unit mass) of protons by low-frequency AW/KAW turbulence in low-$\beta$ plasmas,

$$Q_{\perp\text{stoch}} = \frac{c_1(\delta v_p)^3}{\rho_p}\exp\left(-\frac{c_2}{\varepsilon_p}\right), \tag{3}$$

where $c_1$ and $c_2$ are dimensionless constants,

$$\varepsilon_p = \frac{\delta v_p}{v_{\perp p}}, \tag{4}$$

$$\delta v_p = \left[\int_{e^{-0.5}\rho_p^{-1}}^{e^{0.5}\rho_p^{-1}} E_v(k_{\perp})dk_{\perp}\right]^{1/2}, \tag{5}$$

and $E_v(k_{\perp})$ is the 1D power spectrum of the $E \times B$ velocity ($cE \times B/B^2$). The normalization of $E_v(k_{\perp})$ is chosen so that $\int_0^{\infty} E_v(k_{\perp})dk_{\perp}$ is the total mean square $E \times B$ velocity. Thus, $\delta v_p$ is the rms amplitude of the $E \times B$ velocity at scale $\rho_p$. Chandran et al. (2010) and Xia et al. (2013) found that in low-$\beta$ plasmas stochastic ion heating increases $T_{\perp}$ much more than it increases $T_{\parallel}$, so that $Q_{\perp\text{stoch}}$ is a perpendicular heating rate. Chandran et al. (2010) numerically simulated stochastic heating of test particles by a spectrum of randomly phased AWs and KAWs and found that $c_1 = 0.75$ and $c_2 = 0.34$. Xia et al. (2013) numerically simulated the stochastic heating of test particles by the time-dependent electromagnetic fields produced by direct numerical simulations of strong, reduced magnetohydrodynamic (RMHD) turbulence. These authors found that the heating rate of the test particles grew as they increased the numerical resolution and breadth of the inertial range in the RMHD simulations. For the largest RMHD simulations that they carried out (with 256 grid points along the direction parallel to the background magnetic field $B_0$ and $1024^2$ grid points in the plane perpendicular to $B_0$), Xia et al.

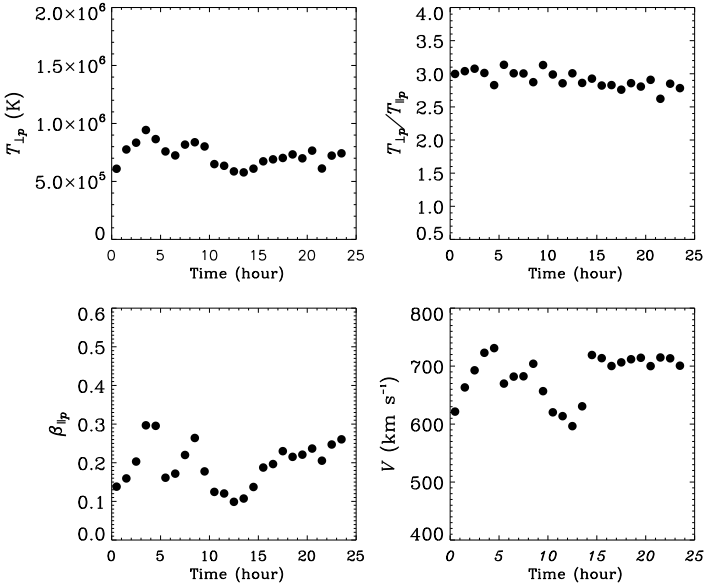

FIG. 1.— Mean values of plasma parameters during each hour of DOY 105 ($r = 0.29$ AU), where $T_{\perp p}$ ($T_{\parallel p}$) is the perpendicular (parallel) proton temperature, $V$ is the proton outflow velocity, and $\beta_{\parallel p}$ is defined in Equation (10).

(2013) found that $c_1 = 0.74$ and $c_2 = 0.21$ for a simulation in which the test-particle gyroradius was approximately equal to the inverse of the dissipation wavenumber in the RMHD simulation (their simulation "D4"). Chandran et al. (2010) and Xia et al. (2013) argued that stochastic heating is more efficient at fixed $\delta v_p$ in strong AW/KAW turbulence than in a randomly phased wave field, implying a larger value of $c_1$ and/or smaller value of $c_2$, because strong AW/KAW turbulence produces coherent structures that increase orbit stochasticity (see also Dmitruk et al. 2004).

The stochastic heating rate $Q_{\perp \text{stoch}}$ can be compared with the perpendicular heating rate per unit mass that is required to explain the non-adiabatic $T_{\perp p}$ profile in the solar wind (Lie-Svendsen et al. 2001; Sharma et al. 2006),

$$Q_{\perp \text{emp}} = BV \frac{d}{dr} \left( \frac{k_B T_{\perp p}}{m_p B} \right), \qquad (6)$$

where $V$ is the radial outflow velocity of the protons. We refer to $Q_{\perp \text{emp}}$ in Equation (6) as the "empirical heating rate." For fast wind with $600 \text{ km/s} \lesssim V \lesssim 700 \text{ km/s}$ at $0.29 \text{ AU} \leq r \leq 0.64$ AU, we take $B$ in Equation (6) to scale as

$$B \propto r^{-n_B}, \qquad (7)$$

where $n_B$ is a constant, and rewrite equation (6) as

$$Q_{\perp \text{emp}} = \frac{k_B V}{m_p r^{n_B}} \frac{d}{dr} \left( r^{n_B} T_{\perp p} \right). \qquad (8)$$

For purely radial magnetic field lines, $n_B$ would equal 2. Because of solar rotation, however, the average magnetic field bends in the azimuthal direction as described by Parker (1958), which causes $B$ to fall off with radius more slowly than $r^{-2}$. We set

$$n_B = 1.55, \qquad (9)$$

which is consistent with the variation in $B$ between the data that we analyze at $r = 0.29$ AU and $r = 0.64$ AU.

## 3. ANALYSIS OF *HELIOS 2* MEASUREMENTS TO EVALUATE $Q_{\perp \text{STOCH}}$ AND $Q_{\perp \text{EMP}}$

In this section, we evaluate $Q_{\perp \text{stoch}}$ and $Q_{\perp p}$ using three days of fast-wind measurements from the *Helios 2* spacecraft: days (DOYs) 76, 95, and 105 from the year 1976. On these days, *Helios 2* was at heliocentric distances of 0.64 AU, 0.40 AU, and 0.29 AU, respectively. We divide each of these days into 24 hour-long intervals and retain only those intervals in which the average value of

$$\beta_{\parallel p} = \frac{8\pi n_p k_B T_{\parallel p}}{B^2} \qquad (10)$$

satisfies

$$\beta_{\parallel p} < 0.3, \qquad (11)$$

where $n_p$ is the proton number density. This selection criterion enables us to test the stochastic-heating theory of Chandran et al. (2010) in the low-$\beta_{\parallel p}$ regime in which Equation (3) was derived. This criterion is also the reason that we do not extend our analysis to larger $r$, since values of $\beta_{\parallel p}$ satisfying Equation (11) become rare and atypical in fast-wind streams at larger $r$. In Section 3.1, we describe our analysis procedure in detail for DOY 105. We then repeat this analysis for DOY 76 and DOY 95 in Section 3.2.

### 3.1. *Proton Heating at $r = 0.29$ AU*

To evaluate $Q_{\perp \text{stoch}}$ and $Q_{\perp \text{emp}}$, we analyze *Helios 2* measurements of the proton velocity distribution function (VDF) and vector magnetic field. The cadence of the plasma experiment on *Helios 2* for obtaining proton VDFs was $\simeq 40$ s (see, e.g., Marsch et al. 1982a). In this work we consider only the proton core (excluding proton beams if present), which represents about 80 percent of the protons (see Bourouaine et al. 2010). By taking moments of the proton-core VDF, we evaluate the proton plasma parameters $n_p$, $T_{\perp p}$, and $T_{\parallel p}$ once every $\simeq 40$ s. The *Helios 2* plasma experiment also reports $\simeq 40$-second averages of the magnetic field vector. We use these 40-s averages to compute $\beta_{\parallel p}$ with a $\simeq 40$-second cadence.

As mentioned above, we divide DOY 105 into twenty-four hour-long intervals. We then compute time-averaged values of $T_{\perp p}$ and $\beta_{\parallel p}$ for each hour. The results are shown in Figure 1. All 24 intervals on DOY 105 satisfy Equation (11). Marsch et al. (1982b) showed that the radial profile of the perpendicular proton temperature follows a power law, $T_{\perp p} \propto r^{-1.08}$, for fast-wind with $600 \text{ km/s} < V < 700 \text{ km/s}$ for the radial interval $0.3 \text{ AU} \leq r \leq 1$ AU. We make the approximation that this average power-law profile for $T_{\perp p}$ applies to the wind streams that we analyze in this paper. The average values of $V$ and $T_{\perp p}$ on DOY 105 are $V = 683$ km/s and $7.24 \times 10^5$ K, respectively. Upon setting $V = 683$ km/s and $7.24 \times 10^5 (r/0.29 \text{ AU})^{-1.08}$ K in Equation (8), we obtain

$$Q_{\perp \text{emp}} = 4.27 \times 10^8 \text{ cm}^2 \text{ s}^{-3} \quad \text{at } r = 0.29 \text{ AU.} \qquad (12)$$

To determine the plasma and turbulence parameters on the right-hand side of Equation (3) at $r = 0.29$ AU, we first evaluate the perpendicular thermal speed $v_{\perp p}$ using the average value of $T_{\perp p}$ on DOY 105, which gives $v_{\perp p} = 109$ km/s . To evaluate $\rho_p$, we use this same value of $v_{\perp p}$ as well as the time average of the magnetic field strength on DOY 105, which is 41.7 nT. These values give a gyroradius of $\rho_p = 27.4$ km.

To evaluate the gyroscale velocity-fluctuation amplitude, we set

$$\delta v_{\rm p} = \frac{\sigma v_{\rm A} \delta B_{\rm p}}{B_0}, \tag{13}$$

where $\sigma$ is a dimensionless constant, and

$$v_{\rm A} = \frac{B_0}{\sqrt{4\pi m_{\rm p}(n_{\rm p} + 4n_\alpha)}} \tag{14}$$

is the Alfvén speed based on the total ion mass (protons plus alphas). In Equation (13), we use the average value of $v_{\rm A}$ during the analyzed hour-long intervals of day 105, which is 155 km/s. The quantity $\delta B_{\rm p}$ is given by

$$\delta B_{\rm p} = \left[ \int_{e^{-0.5}\rho_{\rm p}^{-1}}^{e^{0.5}\rho_{\rm p}^{-1}} E_B(k_\perp) dk_\perp \right]^{1/2}, \tag{15}$$

where $E_B(k_\perp)$ is the 1D wavenumber spectrum of the magnetic field in the plasma rest frame. The normalization of $E_B(k_\perp)$ is chosen so that $\int_0^\infty E_B(k_\perp) dk_\perp$ is the mean of the square of the vector magnetic field fluctuation. For AWs with $k_\perp \rho_{\rm p} \ll 1$, $\sigma = 1$. On the other hand, for linear KAWs with $k_\perp \rho_{\rm p} \gtrsim 1$, $\sigma > 1$. We adopt the value

$$\sigma = 1.19, \tag{16}$$

which was obtained by Chandran et al. (2010) for a spectrum of randomly phased kinetic Alfvén waves with $k_\perp \rho_{\rm p} \simeq 1$ and $k_\parallel$ values chosen in accord with critical-balance models (Goldreich & Sridhar 1995; Cho & Lazarian 2004). (In Figure 6 below, we illustrate how our results change as we vary $\sigma$ between 1.0 and 1.38.)

To evaluate $\delta B_{\rm p}$, we use data from the *Helios 2* flux-gate magnetometer, which had a sampling frequency of 4 Hz for the days we analyze in this paper, corresponding to a Nyquist frequency of 2 Hz. For each of the hour-long intervals of day 105, we compute the frequency spectrum of the magnetic field $P_f(f)$, as described in detail in Appendix A. We plot one such frequency spectrum, for the time interval 00:00:00 - 01:00:00 on day 105, in Figure 2. To relate $P_f(f)$ to the magnetic power spectrum in the plasma rest frame, we treat the solar wind as an infinite homogeneous plasma flowing past the spacecraft at a constant velocity $\boldsymbol{V}$. In accord with Taylor's hypothesis (Taylor 1938), we treat the magnetic field as a time-independent function of space in the plasma rest frame, so that the time variations seen by *Helios 2* result solely from the advection of spatial structure past the spacecraft. Although the time interval $T$ for each FFT is one hour, we treat $T$ as effectively infinite. As shown in Appendix A, these approximations lead to the relation

$$P_f(f) = 2\pi \int P_{\rm 3D}(\boldsymbol{k})\delta(\boldsymbol{k}\cdot\boldsymbol{V} - 2\pi f)d^3 k, \tag{17}$$

where the $k$ integration is over all of $k$-space. The same formula appears in the studies by Horbury et al. (2008) and Forman et al. (2011), but without the factor of $2\pi$ in front of the integral on the right-hand side, implying a different normalization/definition of $P_f(f)$. For our normalization, upon integrating Equation (17) over frequency, we immediately find that $\int_{-\infty}^\infty P_f(f) df = \int d^3 k P_{\rm 3D}(\boldsymbol{k})$. Also, as shown in Appendix A, $\int_{-\infty}^\infty P_f(f) df$ is equal to the average of the square of the fluctuating magnetic field vector.

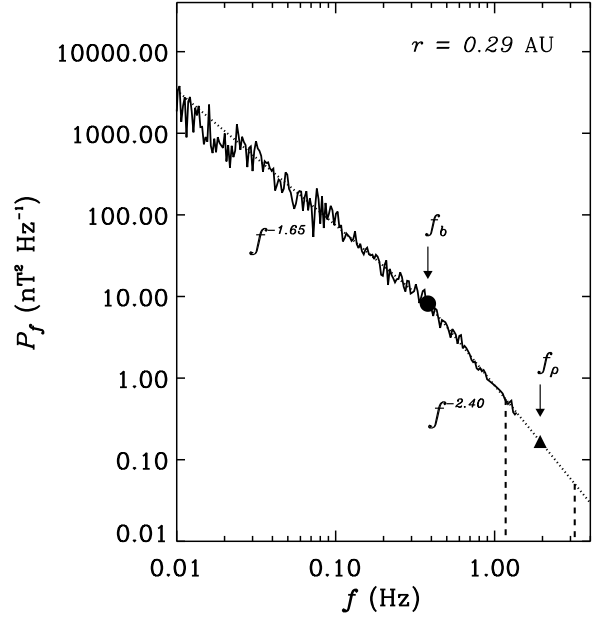

FIG. 2.— Solid line shows the magnetic power spectrum $P_f$ in a fast solar wind stream at $r = 0.29$ AU for the one-hour interval 00:00 to 01:00 of DOY 105 of 1976. The filled circle corresponds to the spectral break frequency, $f_b$. The dotted line is power-law fit to $P_f(f)$. The triangle corresponds to the Doppler-shifted proton frequency $f_\rho$ defined in Equation (20). The rms amplitude of the gyroscale magnetic fluctuation $\delta B_\rho$ is calculated by integrating the power-law fit to $P_f$ between the vertical red-dashed lines.

We assume that the turbulence is "quasi-2D," so that fluctuations vary rapidly in directions perpendicular to the background magnetic field and slowly along the background magnetic field $\boldsymbol{B}_{\rm bg}$. If $\boldsymbol{B}_{\rm bg}$ were uniform in space, then the wavevectors of the magnetic fluctuations would be nearly orthogonal to $\boldsymbol{B}_{\rm bg}$. However, numerical simulations have shown that anisotropic structures in incompressible MHD turbulence are aligned with the local magnetic field rather than the large-volume average of the magnetic field, implying that $\boldsymbol{B}_{\rm bg}$ varies in space (Cho & Vishniac 2000). For a spatially varying $\boldsymbol{B}_{\rm bg}$, it is not entirely clear how to connect the quasi-2D assumption to the mathematical form of $P_{\rm 3D}(\boldsymbol{k})$ in Equation (17) (see, e.g., Matthaeus et al. 2012). Our approach to this problem is similar to the approach employed by Horbury et al. (2008) and Forman et al. (2011). We take $P_{\rm 3D}(\boldsymbol{k})$ in Equation (17) to be cylindrically symmetric about a fixed direction $\hat{\boldsymbol{b}}$ and define the angle between $\hat{\boldsymbol{b}}$ and $\boldsymbol{V}$ (which we treat as constant) to be $\theta_{VB}$. We thus set $P_{\rm 3D}(\boldsymbol{k}) = P_{\rm 3D}(k_\perp, k_\parallel)$, where $k_\parallel$ ($k_\perp$) is the component of $\boldsymbol{k}$ parallel (perpendicular) to $\hat{\boldsymbol{b}}$. We then implement the quasi-2D assumption by taking $P_{\rm 3D}(k_\perp, k_\parallel)$ to be negligible unless $|k_\parallel| \ll k_\perp$. The 1D spectrum $E_B(k_\perp)$ is related to $P_{\rm 3D}(k_\perp)$, through the equation

$$E_B(k_\perp) = 2\pi \int_{-\infty}^\infty P_{\rm 3D}(k_\perp, k_\parallel) k_\perp dk_\parallel. \tag{18}$$

To determine $\theta_{VB}$, we calculate an angle $\psi_i$ once every $\simeq 40$ s using the $\simeq 40$ s values of $\boldsymbol{V}$ and $\boldsymbol{B}$ from the plasma experiment on *Helios 2*. (The $\simeq 40$-s vector magnetic field reported by the plasma experiment is the time average of the magnetic-field vectors measured by the flux-gate magnetome-ter during that $\simeq 40$-s interval.) Within each hour, we have $N \simeq 90$ such measurements. We then set

$$\theta_{VB} = \left( \frac{1}{N} \sum_{i=1}^{N} \psi_i^2 \right)^{1/2}, \qquad (19)$$

so that $\theta_{VB}$ is the rms of the $\simeq 40$-s angles.

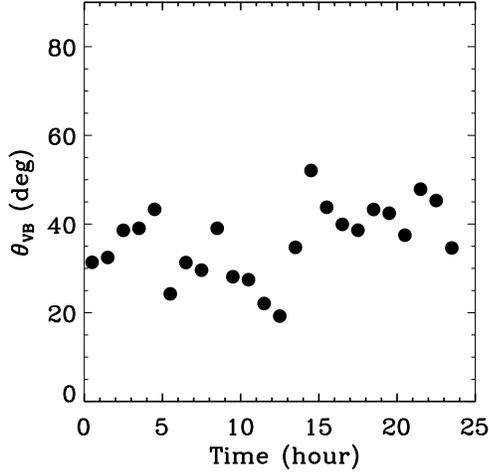

FIG. 3.— Rms values of the angle $\theta_{VB}$ between the $\simeq 40 - s$ measurements of the proton velocity $\boldsymbol{V}$ and magnetic field $\boldsymbol{B}$ during each one-hour time interval of DOY 105.

Spacecraft measurements at $r \simeq 1$ AU find that $P_f(f)$ typically has one power-law slope extending from $\sim 10^{-3}\,\mathrm{s}^{-1}$ to $\sim 1\,\mathrm{s}^{-1}$ and a second, steeper power law extending from $\sim 1\,\mathrm{s}^{-1}$ to $\sim 40\,\mathrm{s}^{-1}$, with either an exponential-like cutoff or an even steeper power law at higher frequencies (Denskat et al. 1983; Bruno & Carbone 2005; Sahraoui et al. 2009; Alexandrova et al. 2011). The *Helios 2* spectra that we analyze from day 105 (as well as DOYs 76 and 95) also show a broken-power-law form, with one power-law scaling at frequencies below a break frequency $f_b$ and a steeper power-law scaling extending from $f_b$ up to the maximum frequency in our spectra, which is $2\,\mathrm{s}^{-1}$. An example of this broken-power-law form can be seen in Figure 2. The numerical values of $f_b$ for each hour of day 105 are shown in Figure 4, along with the frequency

$$f_\rho = \frac{V \sin(\theta_{VB})}{2\pi\rho_p}, \qquad (20)$$

which is the inverse of the time required for a field-aligned structure of transverse dimension $\rho_p$ to be advected past the spacecraft at (vector) velocity $\boldsymbol{V}$. (See Bourouaine et al. (2012) for a discussion of the radial evolution of $f_b$.) The values of $f_\rho$ are shown in Figure 4.[2] As this figure shows,

$$f_b < e^{-0.5} f_\rho, \qquad (21)$$

so that the frequency $e^{-0.5} f_\rho$ is in the part of the spectrum corresponding to the steeper power law. Motivated by the 1-AU measurements described above, we take the power-law scaling at $f_b < f < 2\,\mathrm{s}^{-1}$ in our spectra to extend to much

higher frequencies and assume that[3]

$$P_f(f) \propto f^{-n_s} \quad \text{for } f > f_b, \qquad (22)$$

where $n_s$ is a constant that we obtain by fitting the data. Given this assumption and Equation (21), we show in Appendix B that

$$\delta B_p = \left[ \frac{1}{C_0} \int_{e^{-0.5} f_\rho}^{e^{0.5} f_\rho} P_f(f) df \right]^{1/2}, \qquad (23)$$

where

$$C_0 = \frac{1}{\pi} \int_0^{\pi/2} (\cos\phi)^{n_s - 1} d\phi. \qquad (24)$$

We compute the values of $\delta B_p$ for each hour-long interval of day 105, and then average these values. We substitute this averaged value of $\delta B_p$ into Equation (13) to evaluate $\delta v_p$, using the time average of $B_0$ (41.7 nT) and $v_A$ (155 km/s) on DOY 105 and the value $\sigma = 1.19$. Finally, we compute $\varepsilon_p$ using the value of $v_{\perp p}$ (109 km/s) corresponding to the average of $T_{\perp p}$ on DOY 105. We list the values of the plasma parameters and turbulence parameters in Table 1. This table includes parameter values for DOY 95 and DOY 76 ($r = 0.40$ AU and $r = 0.64$ AU, respectively), but we postpone a discussion of these values until Section 3.2.

To investigate whether stochastic heating can explain the perpendicular proton heating that is occuring in the fast solar wind streams that we analyze, we set

$$Q_{\perp\mathrm{stoch}} = Q_{\perp\mathrm{emp}} \qquad (25)$$

using the values of $\delta v_p$ and $\varepsilon_p$ in Table 1. There are two unknowns in Equation (25): the two dimensionless constants $c_1$ and $c_2$ in the analytical formula for the stochastic heating rate, Equation (3). The solution to Equation (25) is thus a curve in the $c_1 - c_2$ plane. We plot this curve in Figure 5.

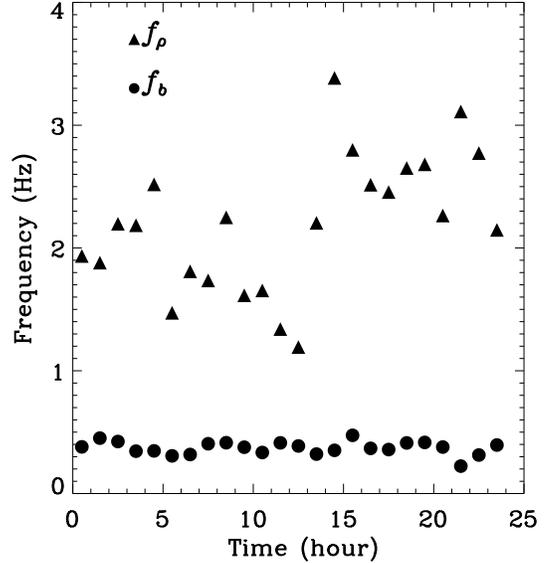

FIG. 4.— Mean values of the frequency $f_\rho$ (upward triangles) and break frequency $f_b$ (circles) in each one-hour time interval of DOY 105. The quantity $f_\rho$ is the inverse of the time required for a field-aligned structure of transverse dimension $\rho_p$ to be advected past the spacecraft at velocity $\boldsymbol{V}$ when the angle $\theta_{VB}$ is computed using Eq. (19).

_______________

[2] Since we calculate $\theta_{VB}$ and $V$ separately for each hour, we use separate one-hour averages of $\rho_p$ and $V$ when evaluating the right-hand side of Equation (20). To determine these one-hour averages of $\rho_p$, we use the one-hour averages of $T_{\perp p}$ and $B$.

[3] To be precise, we assume that $E_B(k_\perp) \propto k_\perp^{-n_s}$ for $k_\perp > e^{-0.5}\rho_p^{-1}$. However, as we show in Appendix B, this latter equation implies that Equation (22) is correct.

| | DOY 105 ($r = 0.29$ AU) | DOY 95 ($r = 0.4$ AU) | DOY 76 ($r = 0.64$ AU) |
|---|---|---|---|
| $V$ (km/s) | 683 | 607 | 601 |
| $\beta_{\parallel p}$ | 0.187 | 0.198 | 0.287 |
| $n_p$ (cm$^{-3}$) | 29.4 | 16.0 | 5.57 |
| $T_{\perp p}$ ($10^5$ K) | 7.24 | 4.36 | 2.70 |
| $B$ (nT) | 41.7 | 24.5 | 12.4 |
| $\theta_{VB}$ | 45.8 | 45.4 | 40.3 |
| $Q_{\perp \mathrm{emp}}$ ($10^8$ cm$^2$ s$^{-3}$) | 4.27 | 1.72 | 0.65 |
| $\delta B_p$ (nT) | 1.16 | 0.70 | 0.32 |
| $\delta v_p$ (km/s) | 5.15 | 4.13 | 3.21 |
| $\varepsilon_p$ | 0.0471 | 0.0486 | 0.0480 |
| $c_2$ | 0.210 | 0.215 | 0.202 |

TABLE 1
AVERAGE QUANTITIES AND REQUIRED STOCHASTIC-HEATING CONSTANT $c_2$ (ASSUMING $c_1 = 0.75$ AND $\sigma = 1.19$).

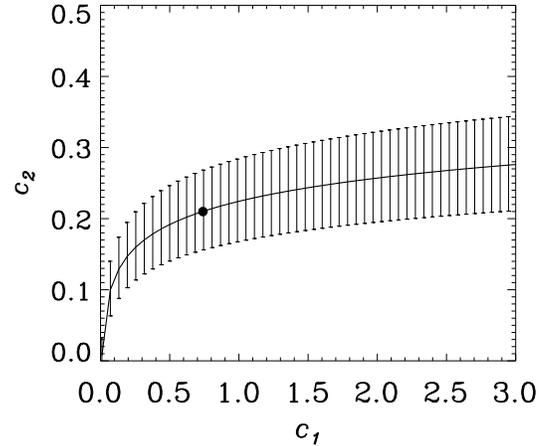

FIG. 5.— Solid-line curve shows the values of $c_1$ and $c_2$ that would be needed in order for the stochastic proton heating rate in Equation (3) to match the empirical proton heating rate in Equation (8) on DOY 105 ($r = 0.29$ AU), assuming $\sigma = 1.19$. The upper (lower) error bar corresponds to $\sigma = 1.38$ ($\sigma = 1$). The filled circle shows the values of $c_1$ and $c_2$ found in a numerical simulation of stochastic heating of test particles by RMHD turbulence carried out by Xia et al. (2013), in which the test-particle gyroradii were comparable to the inverse of the dissipation wavenumber (their Simulation D4).

Because $c_2$ appears in the exponential in the formula for $Q_{\perp \mathrm{stoch}}$, it has a much larger effect on $Q_{\perp \mathrm{stoch}}$ than does $c_1$. It is thus useful to consider the value of $c_2$ for which Equation (25) is satisfied for a fixed, plausible choice of $c_1$. As discussed in Section 2, Chandran et al. (2010) found that $c_1 = 0.75$ for test-particles heated by randomly phased AWs and KAWs, and Xia et al. (2013) found that $c_1 = 0.74$ for test particles heated by strong RMHD turbulence when the particle gyroradii are comparable to the inverse of the dissipation wavenumber of the turbulence (their Simulation D4). We list in Table 1 the value of $c_2$ for which $Q_{\perp \mathrm{stoch}} = Q_{\perp \mathrm{emp}}$ when $c_1 = 0.75$. The filled circle in Figure 5 shows the values of $c_1$ and $c_2$ in Simulation D4 of Xia et al. (2013).

We now consider how the values of $\varepsilon_p$ and $c_2$ in the $r = 0.29$ AU column of Table 1 would change if we were to alter some of the assumptions made in our analysis. If we were to set $\sigma = 1$ instead of $\sigma = 1.19$, then $\varepsilon$ would decrease by 17% and $c_2$ would decrease by 26%. On the other hand, if we were to set $\sigma = 1.38$ instead of $\sigma = 1.19$, then $\varepsilon$ would increase by 15% and $c_2$ would increase by 27%. If we were to use one-hour averages of $V$ and $B$ to determine $\theta_{VB}$ instead of 40-second averages, then $\theta_{VB}$ would decrease to values between 10 and 15 degrees, $f_p$ would be smaller, $\varepsilon$ would increase by approximately 50%, and $c_2$ would approximately double. However, as discussed previously, Cho & Vishniac (2000) showed that small-scale structures are aligned with the local magnetic field direction rather than with the direction of a long-time average of the magnetic field, and so we do not consider these smaller values of $\theta_{VB}$ to be realistic. If we were to ignore the alpha-particle contribution to the mass density when calculating the Alfvén speed, we would infer a larger value of $\delta v_p$, and $c_2$ would increase by approximately 10%. If we were to set $B \propto r^{-2}$ in Equation (6) instead of $B \propto r^{-1.55}$, then $Q_{\perp \mathrm{emp}}$ would increase by 95% and $c_2$ would decrease by 15%. The above discussion illustrates that the most important source of uncertainty in our analysis is the value of $\sigma$. We include the values of $c_2$ corresponding to $\sigma = 1$ and $\sigma = 1.38$ as error bars around our $\sigma = 1.19$ results in Figures 5 and 6.

### 3.2. Proton Heating at $0.29$ AU $\leq r \leq 0.64$ AU

In this section, we repeat the analysis of Section 3.1 for days 76 and 95 of the year 1976. On these days, the *Helios 2* spacecraft was at $r = 0.64$ AU and $r = 0.40$ AU, respectively. On day 76 (95), only six (eight) of the twenty-four hour-long intervals satisfy our selection criterion $\beta_{\parallel p} < 0.3$. We restrict our analysis to the selected hours, so that measurements from intervals with $\beta \geq 0.3$ do not enter into the calculation in any way. For each hour-long interval that we analyze, Equation (21) is satisfied, and so we employ Equations (23) and (24) to evaluate $\delta B_p$.

The plasma parameters and turbulence parameters resulting from our analysis are listed in Table 1. The values of $Q_{\perp \mathrm{emp}}$, $T_{\perp p}$, $n_p$, $B$, $\delta v_p$, and $\delta B_p$ all decrease significantly as $r$ increases from $0.29$ AU to $0.64$ AU. However, the value of $\varepsilon_p$ is almost independent of $r$. In addition, the value of $c_2$ needed to satisfy Equation (25) when $c_1 = 0.75$ is almost independent of $r$. It is unlikely that true values of $c_2$ that characterize stochastic heating in the solar wind vary by a large factor between different low-$\beta$ fast-wind streams, because the physical conditions and turbulence properties are similar. Thus, if, contrary to fact, the three values of $c_2$ in Table 1 varied by a factor of 2, then this variation would be inconsistent with the hypothesis that stochastic heating is the dominant perpendicular heating mechanism for protons in low-$\beta_{\parallel p}$ fast-solar-wind streams. On the other hand, the approximate invariance of $c_2$ in Table 1 as $r$ increases from $0.29$ AU to $0.64$ AU is consistent with this hypothesis.

### 4. CONCLUSIONS

In this paper, we use *Helios 2* measurements to investigate whether stochastic heating by low-frequency AW/KAW turbulence can provide an explanation for the perpendicular proton heating that occurs in low-$\beta$ fast-solar-wind streams at heliocentric distances $0.29$ AU $\leq r \leq 0.64$ AU. We discuss in detail our method for inferring the plasma parameters and turbulence properties from the measurements on day 105 of year 1976, when *Helios 2* was at $r = 0.29$ AU. We then repeat this analysis for days 76 and 95 of year 1976, on which the satellite was at $r = 0.64$ AU and $r = 0.40$ AU, respectively. We restrict our analysis to hour-long intervals on each day in which the average value of $\beta_{\parallel p}$ is $< 0.3$. We use measurements of the mean values of the proton perpendicular temperature $T_{\perp p}$ and outflow velocity $V$ to infer empirically the value of the perpendicular proton heating rate $Q_{\perp\text{emp}}$. We then analyze *Helios-2* magnetometer data in the low-$\beta_{\parallel p}$ data to obtain the amplitude of turbulent fluctuations at the proton-gyroradius scale. We use this turbulence amplitude to evaluate the analytic formula derived by Chandran et al. (2010) for the stochastic heating rate $Q_{\perp\text{stoch}}$. By setting $Q_{\perp\text{stoch}} = Q_{\perp\text{emp}}$, we obtain a constraint on the two dimensionless constants $c_1$ and $c_2$ that appear in the analytical formula for $Q_{\perp\text{stoch}}$.

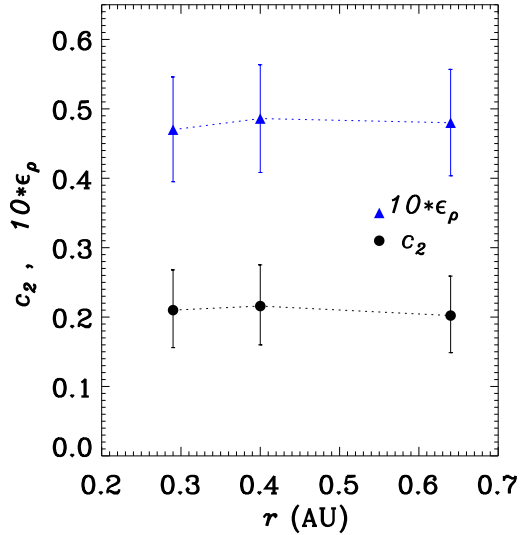

FIG. 6.— Triangles show the parameter $\epsilon_p$ as a function of heliocentric distance $r$, assuming $\sigma = 1.19$. Circles show the value of the stochastic-heating constant $c_2$ that would be needed in order for $Q_{\perp\text{stoch}}$ to equal $Q_{\perp\text{emp}}$, assuming $\sigma = 1.19$. Upper and lower error bars correspond, respectively, to $\sigma = 1.38$ and $\sigma = 1$.

Our results are consistent with the hypothesis that stochastic heating by low-frequency AW/KAW turbulence accounts for much of the perpendicular proton heating that occurs in low-$\beta_{\parallel p}$ fast solar wind streams, for two reasons. First, the constraint that we derive on the two constants $c_1$ and $c_2$ is consistent with the numerical values of $c_1$ and $c_2$ that were obtained by Xia et al. (2013) in a numerical simulation of stochastic heating of test particles by strong RMHD turbulence in which the test-particle gyroradii were comparable to the inverse of the dissipation wavenumber of the turbulence. Second, the required value of the constant $c_2$ (assuming $c_1$ is of order unity) is almost independent of $r$.

The main source of uncertainty in our analysis lies in the value of the dimensionless constant $\sigma$ appearing in Equation (13), which relates the amplitude of the gyroscale velocity fluctuations to the amplitude of the gyroscale magnetic-field fluctuations. We have adopted the value $\sigma = 1.19$, which was obtained by Chandran et al. (2010) in a model of critically balanced, randomly phased AWs/KAWs. However, the precise value of $\sigma$ in the solar wind is not known. In Figure 6 we show how our constraint on the constant $c_2$ changes as we vary $\sigma$ from 1 to 1.38. Another source of error in our analysis is our extrapolation of the observed power law in $P_f(f)$ at $f_b < f < 2\,\text{s}^{-1}$ to frequencies $> 2\,\text{s}^{-1}$. At sufficiently high frequencies, this power spectrum must steepen due to strong dissipation. By neglecting such steepening, we overestimate $\delta B_p$ to some degree. The values of the stochastic-heating constants $c_1$ and $c_2$ obtained in numerical simulations may also differ from the values that describe stochastic heating in the solar wind. The test-particle/RMHD simulations of Xia et al. (2013) differ from solar-wind turbulence in that the inertial range in the simulations is not as broad as in the solar wind, the RMHD code does not account for KAW physics (proton/electron decoupling) at $k_\perp \rho_p \gtrsim 1$, and the process of stochastic proton heating does not have a back reaction upon the turbulent fields. Future work to provide more realistic simulations of stochastic heating and to reduce the uncertainties in $\sigma$ and $\delta B_p$ would lead to a more rigorous test of the importance of stochastic heating in the solar wind.

We thank the anonymous referee for comments and suggestions that helped improve the manuscript. This work was supported in part by grant NNX11AJ37G from NASA's Heliophysics Theory Program, NASA grant NNX12AB27G, NSF/DOE grant AGS-1003451, DOE grant DE-FG02-07-ER46372, and NSF grant AGS-0851005.

## APPENDIX

### A. CONNECTING THE MEASURED FREQUENCY SPECTRUM OF THE MAGNETIC FIELD $P_F(F)$ TO THE 3D WAVENUMBER SPECTRUM $P_{3D}(K)$

We treat the solar wind as an infinite, statistically homogeneous medium that flows past the spacecraft at a constant velocity $\boldsymbol{V}$. On the basis of Taylor's hypothesis (Taylor 1938), we treat the magnetic field in the solar-wind frame as a time-independent function of position $\boldsymbol{B}(\boldsymbol{x})$. We define the auto-correlation function of the magnetic field to be

$$h(\boldsymbol{x}) = \langle \boldsymbol{B}(\boldsymbol{x}_0) \cdot \boldsymbol{B}(\boldsymbol{x}_0 + \boldsymbol{x}) \rangle, \tag{A1}$$

where $\langle \ldots \rangle$ denotes an average over an ensemble of realizations of the turbulence, which we assume is equivalent to an average over $\boldsymbol{x}_0$. We define the 3D magnetic power spectrum to be

$$P_{3D}(\boldsymbol{k}) = \frac{h_{\boldsymbol{k}}}{(2\pi)^3}, \tag{A2}$$

where

$$h_{\boldsymbol{k}} = \int h(\boldsymbol{x}) e^{-i\boldsymbol{k}\cdot\boldsymbol{x}} d^3 x \tag{A3}$$

is the Fourier transform of $h(\boldsymbol{x})$, which satisfies the inverse transform relation

$$h(\boldsymbol{x}) = \frac{1}{(2\pi)^3} \int h_{\boldsymbol{k}} e^{i\boldsymbol{k}\cdot\boldsymbol{x}} d^3k. \tag{A4}$$

Upon integrating Equation (A2) over $k$, we obtain the relation

$$\int P_{3\mathrm{D}}(\boldsymbol{k}) d^3k = \langle |\boldsymbol{B}(\boldsymbol{x}_0)|^2 \rangle. \tag{A5}$$

Since the solar wind flows past the spacecraft at velocity $\boldsymbol{V}$, the velocity of the spacecraft in the solar-wind frame is $-\boldsymbol{V}$. The magnetic field measured by the spacecraft can thus be written

$$\boldsymbol{H}(t) = \boldsymbol{B}(\boldsymbol{x}_0 - \boldsymbol{V}t), \tag{A6}$$

where $\boldsymbol{x}_0$ is the position vector of the spacecraft in the solar-wind frame at $t = 0$. We define the temporal autocorrelation function

$$R(t) = \langle \boldsymbol{H}(t_0) \cdot \boldsymbol{H}(t_0 + t) \rangle, \tag{A7}$$

where $\langle \ldots \rangle$ again refers to an average over an ensemble of realizations of the turbulence, which we now assume is equivalent to an average over $t_0$. We define the frequency power spectrum

$$P_f(f) = \tilde{R}(f), \tag{A8}$$

where

$$\tilde{R}(f) = \int_{-\infty}^{\infty} R(t) e^{2\pi i f t} dt \tag{A9}$$

is the temporal Fourier transform of $R(t)$. As is common in studies of solar-wind turbulence, we work with the frequency $f$ instead of the angular frequency $\omega = 2\pi f$. In contrast to the right-hand side of Equation (A2), which includes a factor of $(2\pi)^{-1}$ for each spatial dimension, we do not include a factor of $(2\pi)^{-1}$ on the right-hand side of Equation (A8), because $P_f(f)$ is the power per unit $f$ rather than per unit $\omega = 2\pi f$. Also, in contrast to Equation (A3), we do not include a minus sign in the exponential in Equation (A9), following the standard convention. It follows from Equations (A1) and (A6) that

$$R(t) = h(-\boldsymbol{V}t). \tag{A10}$$

Upon substituting Equation (A10) into Equation (A8) and carrying out the integration over $t$, we find that

$$P_f(f) = 2\pi \int P_{3D}(\boldsymbol{k}) \delta(2\pi f - \boldsymbol{k}\cdot\boldsymbol{V}) d^3k. \tag{A11}$$

Upon integrating Equation (A11) over $f$, we obtain

$$\int_{-\infty}^{\infty} P_f(f) df = \int P_{3D}(\boldsymbol{k}) d^3k, \tag{A12}$$

where the $k$ integration is over all of $k$ space. It then follows from Equation (A5) that

$$\int_{-\infty}^{\infty} P_f(f) df = \langle |\boldsymbol{B}(\boldsymbol{x}_0)|^2 \rangle. \tag{A13}$$

For the purposes of analyzing spacecraft data, it is useful to relate $P_f(f)$ to the Fourier transform of the magnetic field. A technical difficulty is that $\boldsymbol{H}(t)$ does not vanish as $t \to \pm\infty$, because we have taken the solar wind to be infinite and statistically homogeneous. To circumvent this difficulty, we define a temporal window function

$$W(t) = \begin{cases} 1 & \text{if } |t| < T/2 \\ 0 & \text{otherwise} \end{cases} \tag{A14}$$

and introduce the function

$$\boldsymbol{J}(t) = W(t)\,\boldsymbol{H}(t). \tag{A15}$$

Because $\boldsymbol{J}(t)$ vanishes for $|t| > T/2$, we can take its Fourier transform, which is given by

$$\tilde{\boldsymbol{J}}(f) = \int_{-\infty}^{\infty} \boldsymbol{J}(t) e^{2\pi i f t} dt. \tag{A16}$$

It follows from Equations (A15) and (A16) that

$$\langle \tilde{\boldsymbol{J}}(f_1) \cdot \tilde{\boldsymbol{J}}(f_2) \rangle = \int_{-T/2}^{T/2} dt_1 \int_{-T/2}^{T/2} dt_2 \, \langle \boldsymbol{H}(t_1) \cdot \boldsymbol{H}(t_2) \rangle e^{2\pi i (f_1 t_1 + f_2 t_2)}. \tag{A17}$$

After changing variables of integration from $(t_1, t_2)$ to $(s, \tau)$, where

$$s = (1/2)(t_1 + t_2) \quad \text{and} \quad \tau = t_2 - t_1, \tag{A18}$$

we obtain

$$\langle \tilde{\boldsymbol{J}}(f_1) \cdot \tilde{\boldsymbol{J}}(f_2) \rangle = \left( \int_{-T/2}^{0} ds \int_{-2s-T}^{2s+T} d\tau + \int_{0}^{T/2} ds \int_{2s-T}^{-2s+T} d\tau \right) R(\tau) e^{2\pi i [(f_1+f_2)s+(f_2-f_1)(\tau/2)]}. \tag{A19}$$

We assume that $R(\tau)$ decays to zero as $|\tau|$ increases above some value $t_{\mathrm{cor}}$, and that $T \gg t_{\mathrm{cor}}$. Thus, to a reasonable approximation (which becomes increasingly accurate as $T/t_{\mathrm{cor}} \to \infty$), we can set

$$\int_{-2s-T}^{2s+T} d\tau \to \int_{-\infty}^{\infty} d\tau \quad \text{and} \quad \int_{2s-T}^{-2s+T} d\tau \to \int_{-\infty}^{\infty} d\tau \tag{A20}$$

in Equation (A19). Substituting Equation (A20) into Equation (A19), we obtain

$$\langle \tilde{\boldsymbol{J}}(f_1) \cdot \tilde{\boldsymbol{J}}(f_2) \rangle \simeq \int_{-T/2}^{T/2} ds \int_{-\infty}^{\infty} d\tau \, R(\tau) e^{2\pi i [(f_1+f_2)s+(f_2-f_1)(\tau/2)]}. \tag{A21}$$

Upon setting $f_1 = -f$ and $f_2 = f$, we find that

$$\langle \tilde{\boldsymbol{J}}(-f) \cdot \tilde{\boldsymbol{J}}(f) \rangle \simeq T \int_{-\infty}^{\infty} R(\tau) e^{2\pi i f \tau} d\tau = T P_f(f) \tag{A22}$$

The fractional errors in Equation (A22) result from the approximations in Equation (A20). These fractional errors vanish as $T \to \infty$. Thus,

$$P_f(f) = \lim_{T \to \infty} \frac{1}{T} \langle \tilde{\boldsymbol{J}}(-f) \cdot \tilde{\boldsymbol{J}}(f) \rangle. \tag{A23}$$

In practice, we do not measure $P_f(f)$ directly, but instead measure a discrete approximation to $P_f(f)$. We follow the discrete-Fourier-transform conventions of Press et al. (1992), which we summarize here. We define

$$\boldsymbol{H}_n = \boldsymbol{B}(\boldsymbol{x}_0 - \boldsymbol{V} t_n), \tag{A24}$$

where $t_n = n\Delta t$, $\Delta t = 0.25$ s is the cadence of the *Helios 2* magnetometer data on day 105, and $n = 0, 1, 2, \ldots N-1$. We take $N$ to be even. We define

$$f_n = \frac{n}{N\Delta t}, \tag{A25}$$

and $\delta f = 1/(N\Delta t)$. The discrete Fourier transform of $\boldsymbol{H}_n$ is then given by

$$\tilde{\boldsymbol{H}}(f_n) = \sum_{j=0}^{N-1} \boldsymbol{H}_j e^{2\pi i f_n t_j} \Delta t. \tag{A26}$$

Our discrete approximation of the frequency power spectrum of the magnetic field is given by

$$P_{\mathrm{sc}}(f_n) = \frac{|\tilde{\boldsymbol{H}}(f_n)|^2}{T}, \tag{A27}$$

where $T = N\Delta t$. This discrete power spectrum satisfies the discrete version of Parseval's theorem,

$$\frac{1}{N} \sum_{n=0}^{N-1} |\boldsymbol{H}_n|^2 = \sum_{n=0}^{N-1} P_{\mathrm{sc}}(f_n) \delta f. \tag{A28}$$

We use $P_{\mathrm{sc}}(f_n)$ instead of $P_f(f)$ when evaluating $(\delta B_{\mathrm{p}})^2$ in Equation (23). We note that $\tilde{\boldsymbol{H}}(f_n)$ is periodic in $f_n$, with $\tilde{\boldsymbol{H}}(f_{n+N}) = \tilde{\boldsymbol{H}}(f_n)$, and thus

$$P_{\mathrm{sc}}(f_{n+N}) = P_{\mathrm{sc}}(f_n). \tag{A29}$$

For even $N$, we can use Equation (A29) to rewrite Equation (A28) as

$$\frac{1}{N} \sum_{n=0}^{N-1} |\boldsymbol{H}_n|^2 = \sum_{n=-N/2+1}^{(N/2)} P_{\mathrm{sc}}(f_n) \delta f. \tag{A30}$$

Because $\boldsymbol{H}_j$ is a real function of time, $[\tilde{\boldsymbol{H}}(f_n)]^* = \tilde{\boldsymbol{H}}(-f_n)$, and $P_{\mathrm{sc}}(-f_n) = P_{\mathrm{sc}}(f_n)$. It is thus sufficient to plot $P_{\mathrm{sc}}(f_n)$ for positive frequencies. However, it is important to note that some numerical packages for computing frequency power spectra compute a "one-sided" power spectrum that is equal to $2P_{\mathrm{sc}}(f_n)$ for $0 < n < N/2$ and equal to $P_{\mathrm{sc}}(f_n)$ for $n = N/2$ and $n = 0$. For such a one-sided power spectrum, the sum on the right-hand side of Equation (A30) is taken from $n = 0$ to $n = N/2$ instead of from $n = -N/2+1$ to $N/2$. We emphasize that the formulas we have derived in this paper are based on the "two-sided" power spectrum defined in Equation (A27).

## B. EVALUATING $\delta B_P$ IN TERMS OF THE MEASURED FREQUENCY SPECTRUM $P_F(F)$

We take $\boldsymbol{B}_{\rm bg}$ to be in the $z$ direction and $\boldsymbol{V}$ to lie in the $xz$-plane. We define $\phi$ to be the angle between $\boldsymbol{k}_\perp$ and the $x$ axis, where $\boldsymbol{k}_\perp = \boldsymbol{k} - k_\parallel \hat{\boldsymbol{z}}$ and $k_\parallel = \hat{\boldsymbol{z}} \cdot \boldsymbol{k}$. As discussed in Section 3, we take $P_{3D}(\boldsymbol{k})$ to be axisymmetric about the $z$ direction and negligible unless $|k_\parallel| \ll k_\perp$. We thus approximate $\boldsymbol{k} \cdot \boldsymbol{V}$ in Equation (A11) as $k_\perp \cdot \boldsymbol{V}$ to obtain

$$P_f(f) = \int_{-\pi}^{\pi} d\phi \int_0^\infty dk_\perp \frac{E_B(k_\perp)}{V_\perp |\cos\phi|} \delta\left(k_\perp - \frac{2\pi f}{V_\perp \cos\phi}\right), \tag{B1}$$

where

$$V_\perp = V \sin(\theta_{VB}) \tag{B2}$$

is the component of $\boldsymbol{V}$ perpendicular to $\boldsymbol{B}_{\rm bg}$, and $E_B(k_\perp)$ is defined in Equation (18). We define $k_{\perp 0} = (2\pi f)/V_\perp$ and $k_\perp^* = k_{\perp 0}/\cos\phi$. For concreteness we take $f > 0$, which implies that the integrand in Equation (B1) is nonzero only for $-\pi/2 < \phi < \pi/2$. Since the integrand is even in $\phi$, we can rewrite Equation (B1) as an integral over the interval $\phi \in (0, \pi/2)$:

$$P_f(f) = 2 \int_0^{\pi/2} \frac{E_B(k_\perp^*)}{V_\perp |\cos\phi|} d\phi \tag{B3}$$

We assume that

$$E_B(k_\perp) = A k_\perp^{-n_s} \quad \text{for } k_\perp > e^{-1/2}\rho_{\rm p}^{-1}. \tag{B4}$$

If $f > e^{-1/2}f_{\rm p}$, where $f_{\rm p}$ is defined in Equation (20), then $k_\perp^* > e^{-1/2}\rho_{\rm p}^{-1}$, and $E_B(k_\perp^*) = A(k_\perp^*)^{-n_s}$ in the integrand of Equation (B3). It thus follows that

$$P_f(f) \propto f^{-n_s} \quad \text{for } f > e^{-1/2}f_{\rm p}. \tag{B5}$$

Because the power-law indices describing $E_B(k_\perp)$ and $P_f(f)$ are the same, we infer the power-law index $n_s$ from the measured $P_f(f)$ at $f > f_{\rm b}$. Upon integrating Equation (B3) from $f = e^{-1/2}f_{\rm p}$ to $f = e^{1/2}f_{\rm p}$, we obtain

$$\int_{e^{-1/2}f_{\rm p}}^{e^{1/2}f_{\rm p}} P_f(f)df = \frac{1}{\pi} \int_{e^{-1/2}\rho_{\rm p}^{-1}}^{e^{1/2}\rho_{\rm p}^{-1}} dk_{\perp 0} \int_0^{\pi/2} d\phi \frac{E_B(k_{\perp 0}/\cos\phi)}{|\cos\phi|}. \tag{B6}$$

Equation (B4) implies that we can set $E_B(k_{\perp 0}/\cos\phi) = (\cos\phi)^{n_s} E_B(k_{\perp 0})$ in the integrand on the right-hand side of Equation (B6). We thus obtain

$$\int_{e^{-1/2}\rho_{\rm p}}^{e^{1/2}f_{\rm p}} P_f(f)df = (\delta B_{\rm p})^2 \left[\frac{1}{\pi}\int_0^{\pi/2}(\cos\phi)^{n_s-1}d\phi\right], \tag{B7}$$

which is equivalent to Equation (23).